\documentclass[12pt]{article}
\usepackage{epsfig,subfigure}
\usepackage{amsmath,amsfonts}

\usepackage{epstopdf}

\def\as{\ensuremath{\alpha_{s}}}
\def\a0{\alpha_0}

\def \nn {\nonumber}

\def\bea {\begin{eqnarray}}
\def\eea {\end{eqnarray}}

\def\be {\begin{equation}}
\def\ee {\end{equation}}
\def\bi {\begin{itemize}}
\def\ei {\end{itemize}}

\ifx\pdfoutput\undefined
\input epsf

\def\figscale#1#2{\epsfxsize=#2\epsfbox{#1.eps}}
\else

\def\figscale#1#2{\pdfximage width#2 {#1.pdf}\pdfrefximage\pdflastximage}
\fi

\begin{document}

\begin{flushright}
YITP-SB-16-4
\end{flushright}

\vskip 0.25 in

\begin{center}
{\Large \bf Yang-Mills Theories at High-Energy Accelerators}

\vskip 0.25 in

George Sterman

\vskip 0.050 in

{\it C.N.\ Yang Institute for Theoretical Physics\\
 and Department of Physics and Astronomy\\ 
Stony Brook University, Stony Brook NY 11794-3840 USA}

\end{center}


\begin{abstract}
I'll begin with a brief review of the triumph of Yang-Mills theory at particle accelerators, a development that began some years after their historic paper.  This story reached a culmination, or at least local extremum, with the discovery at the Large Hadron Collider of a Higgs-like scalar boson in 2012.   The talk then proceeds to a slightly more technical level, discussing how we derive predictions from the gauge field theories of the Standard Model and its extensions for use at high energy accelerators.   
\end{abstract}



\section{The Triumph of Yang-Mills Theory at Accelerators}

 High energy accelerators offer the most direct window to short-lived quantum processes.   The strategy of probing matter at short distances has resulted in the identification/discovery of the gauge and matter fields of the Standard Model.   Accelerator programs, however complex and costly, remain experiments that follow the scientific canon. They are capable of design, replication and variation in response to the demands of nature and the imagination.   The series of accelerator-based experiments of the past fifty years led ineluctably to the triumph of the gauge concept \cite{Yang:1954ek}.

\begin{figure}
\centerline{\figscale{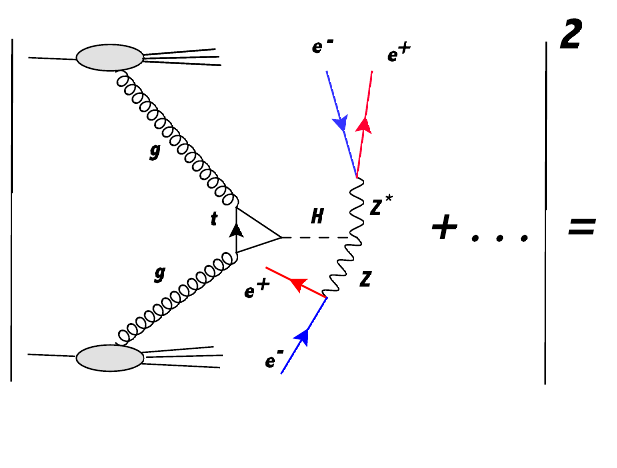}{6cm} \quad \figscale{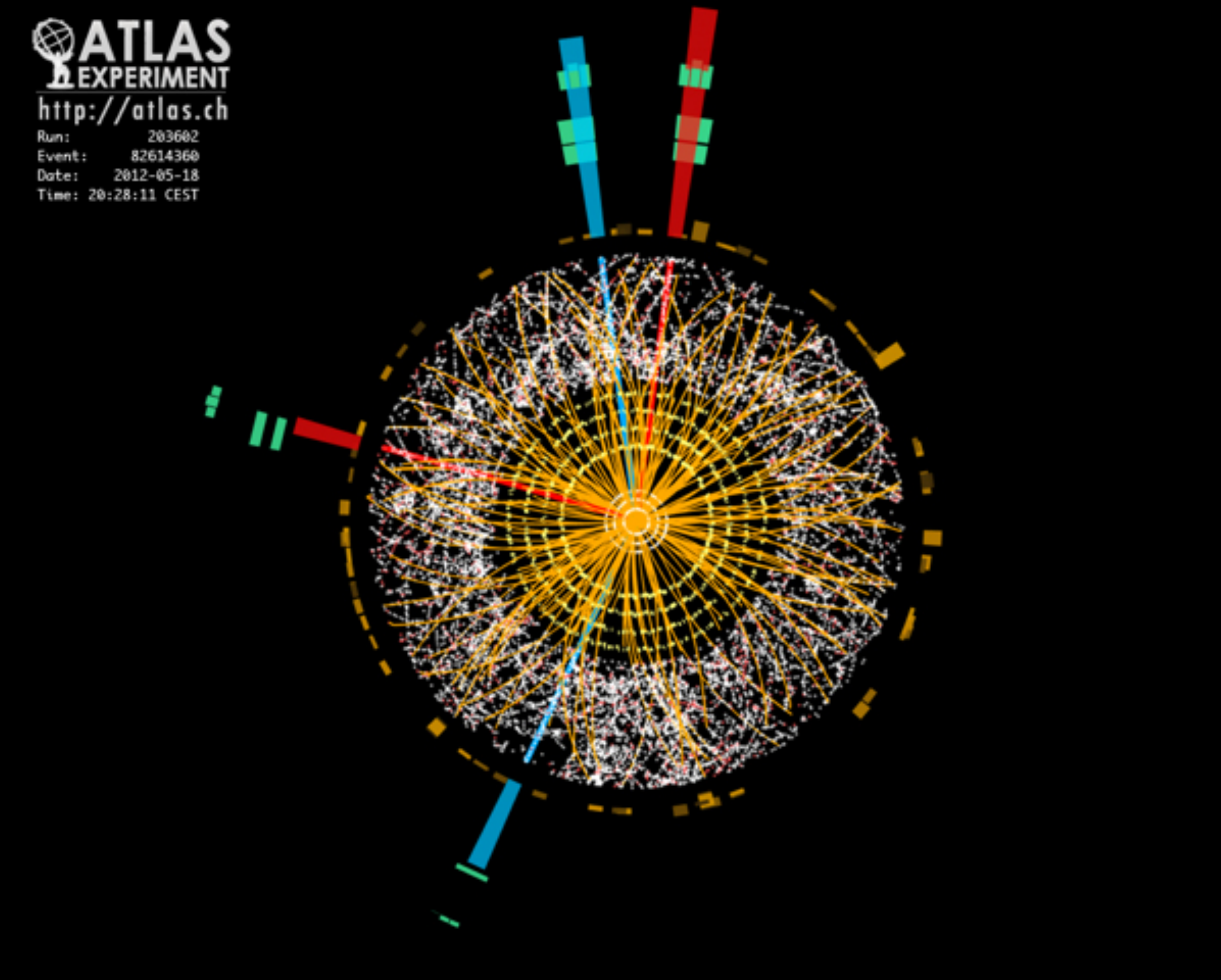}{6cm}}
\caption{Schematic equivalence between the squared microscopic amplitude and the ATLAS experiment detector signal for Higgs production followed by decay to four leptons through $Z$ bosons.}
\label{fig:higgs-ZZ}
\end{figure}

 I will review a little of how quantum field theory is applied in accelerator experiments, but we can sum it up with Fig.\ \ref{fig:higgs-ZZ}, a picture worth a thousand words.

On the left, the figure shows a quantum-mechanical history that includes all the essential elements of the spontaneously-broken gauge theories \cite{Englert:1964et,Higgs:1964pj,Guralnik:1964eu}   of the Standard Model \cite{Glashow:1961tr}-\cite{Weinberg:1973un}.   Once it became clear that such theories lend themselves to quantum mechanical renormalization \cite{'tHooft:1972fi}, and with the discovery of asymptotic freedom \cite{Gross:1973id,Politzer:1973fx}, it was possible to create systematic methods to propose, predict and test for experimental signatures that reflect directly their fundamental structure.   The left of Fig. \ref{fig:higgs-ZZ} shows the merger of $SU(3)$-colored gluons through a top quark loop, followed by  the production of a Higgs boson \cite{Aad:2012tfa,Chatrchyan:2012xdj}, linking the top quarks to the $SU(2)_L\times U(1)$ electroweak sector of the Standard Model, with the transient appearance of Z bosons, and their subsequent decay into lepton pairs.

 In quantum field theory, every observed final state is the result of a quantum-mechanical set of stories of this type.  So far, the stories supplied by the Standard Model, built on an unbroken $SU(3)$ color gauge theory (very much like the original Yang-Mills Lagrangian) with gluons the gauge bosons, and a spontaneously-broken $SU(2)_L\times U(1)$, with $W^\pm,\, Z$ and photons, account for and explain essentially all observations at accelerators.   The gluons and electroweak bosons in this process are themselves gauge bosons,  whose identities and self-interactions disclose the underlying group structure \cite{Yang:1954ek} of the Standard Model.   All other particles observed at colliders, with a variety and range of masses that remain mysterious, are seen precisely because of their gauge-theory interactions.   Indeed, without these interactions we would have no way to produce them at all.
 
  The signature feature of the Yang-Mills extension of gauge invariance to nonabelian groups is the interaction of gauge fields among themselves.   This, of course, is a direct consequence of the form of the gauge field strength, given in the notation of Ref.\ \cite{Yang:1954ek} as
 \bea
 F_{\mu\nu}\ =\ \frac{\partial B_\mu}{\partial x^\nu}\ -\ \frac{\partial B_\nu}{\partial x^\mu}\ +\ i\epsilon \left( B_\mu B_\nu\ -\ B_\nu B_\mu\right)\, ,
 \label{eq:ym-fs}
 \eea
 in terms of gauge fields expressed as matrices.   In ``pure-Yang-Mills" the classical Lagrange density is just ${\cal L}=-(1/4)F^{\mu\nu}F_{\mu\nu}$, invariant under local group (gauge) transformations.     The quadratic terms of the field strength are necessary for this invariance, so that the interactions among the gauge bosons, and with other fields ``are essentially determined by the requirement of gauge invariance" \cite{Yang:1954ek}.   
 
\begin{figure}
 \centerline{\figscale{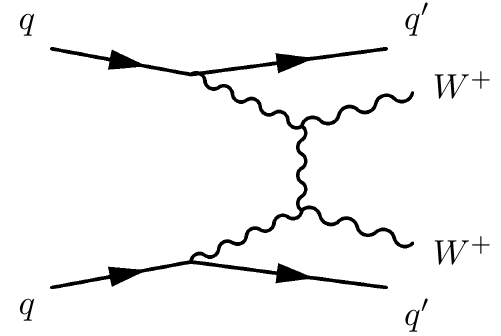}{6cm}}
 \caption{An example of electroweak vector boson scattering at a hadron collider, from Ref.\ \cite{Aad:2014zda}. \label{fig:ww}}
\end{figure}

The self-interactions of the $SU(3)$ component of the Standard Model, quantum chromodynamics (QCD) are the origin of asymptotic freedom, and also contribute directly to nearly every multijet cross section, as described below.    The self couplings of the electroweak gauge bosons also hew closely to the Standard Model (for a review, see Chap.\ 10 of Ref.\ \cite{Agashe:2014kda}), and their study at high energies continues at the Large Hadron Collider \cite{Khachatryan:2015sga,Aad:2014mda}.   At LHC energies  the mutual scattering of electroweak bosons, through processes like those shown in Fig.\ \ref{fig:ww}, is for the first time coming into focus \cite{Khachatryan:2014sta,Aad:2014zda}.

 All this could be the ``end of the story", except that: 1)  Cosmological observations strongly suggest that there are other sources of gravitation in the universe: dark matter, dark energy.  Dark matter is by definition bereft of at least electromagnetic interactions.    2)  The mass of the Higgs particle in the Standard Model in isolation is unstable to overwhelming quantum corrections, a conundrum often referred to as the ``hierarchy" problem.   If the existence of as-yet unobserved particles resolves this problem, they may or may not participate in the gauge interactions of the Standard Model.
 
Contemporary distress with the hierarchy problem of the Standard Model may be compared to 17th Century objections to action at a distance in Newtonian gravity.   The objection comes from profound intuition, but does not immediately suggest a resolution.  It is attractive to suggest that dark matter plays a role, although this is just a guess.     Putting all this aside, as the progress of science put gravitational action at a distance aside until 1915, the success of the gauge-theory based Standard Model is extraordinary.   And resolutions of the Standard Model's puzzles, and even of dark matter and energy, may in the fulness of time come from theories with many or most of the Standard Model's properties, or from generalizations inspired by it, like supersymmetry.  Hopefully, we won't have to wait as long!   For the remainder of this talk, I'll try to explain how accelerator studies helped us get to this stage, how we learned to recount  and recognize the stories like those of Fig.\ \ref{fig:higgs-ZZ} that led to the Standard Model's successes.  

\section{Techniques from Quantum Field Theory}

High energy collisions make possible large momentum transfers, and correspondingly processes that develop over very short distances and times.   As we shall see, for  short distances accessible to accelerators, we can expand around the free field theory.   Starting with an initial state, the system evolves via transitions through one or more intermediate states, finally ending up in an observed final state.     The list of possible transitions between states {\it are} the stories that provide predictions through the calculation of quantum mechanical amplitudes.   The systematic computation of amplitudes in this way is perturbation theory.   A related discussion of the following can be found in Ref.\ \cite{Sterman:2000dj}.

\subsection{Perturbation theory}

 Perturbation theory really just follows from the Schr\"odinger equation, describing the mixing of free particle states (more on this later),
\bea
i\hbar\, {\partial \over \partial t}|\psi(t)>\ =\ \left( H^{(0)}+V\right)|\psi(t)>\, ,
\label{eq:Schrodinger}
\eea
with $H^{(0)}$ the ``free Hamiltonian", and $V$ some potential.  The form of the free Hamiltonian determines the list of possible free states.    Usually we start with free-state ``in"  boundary conditions,
\bea
|\psi(t=-\infty)>\ =\ |m_0>\  =\  | p_1^{\rm in}, p_2^{\rm in}\rangle \, ,
\label{eq:in-state}
\eea
corresponding to two particles that approach each other from the distant past, as prepared in an accelerator.

 Theories differ in their lists of particles and their (hermitian) potentials, sets of operators represented by $V$.   The expansions of perturbation theory are given in terms of matrix elements of $V$ between free states, for which we adopt the notation,
 \bea
V_{j\leftarrow i}=\langle m_j|V|m_i\rangle\, .
\label{eq:potential-element}
\eea
These matrix elements are represented graphically by vertices in diagrams like those on the left of Fig.\ \ref{fig:higgs-ZZ}.  The states $|m_i\rangle$ and $|m_j\rangle$ differ by the absorption or emission of a particle, the annihilation (creation) of a pair  of particles into (from) a heavier particle, etc.  In gauge theories, the difference between the numbers of particles is limited to three.  Many particles can be created, but only by repeated actions of operator $V$.    

Solutions to the Schr\"odinger equation (\ref{eq:Schrodinger}) are sums of  ordered time integrals over the matrix elements in (\ref{eq:potential-element}).   The result is often referred to as ``old-fashioned perturbation theory".    The scattering amplitudes computed this way are precisely equivalent to the results of computing with the more familiar Feynman diagrams.   Scattering experiments measure the quantum mechanical overlap between a state prepared far in the past (an ``in" state, as in Eq.\ (\ref{eq:in-state})) and a state observed far in the future (an ``out" state).   In states are what high energy accelerators like the LHC provide; out states are what detectors like ATLAS and CMS detect.   

It's not difficult to verify \cite{Sterman:2000dj} that, in the notation of Eq.\ (\ref{eq:potential-element}) the overlap between a solution to Eq.\ (\ref{eq:Schrodinger}) with initial condition $|m_0\rangle$ at $t=-\infty$ and a ``final state" $|m_{\rm out}\rangle$ can be written in as
\bea
\langle m_{\rm out}(\infty)| m_0\rangle &=& \sum_{n=0}^\infty\ \sum_{m_1\dots m_{n-1}} 
\ \ \prod_{a=0}^{n} \left( \frac{-i}{\hbar}\ V_{a \leftarrow a-1} \right )\ \int_{-\infty}^\infty d\tau_n \int_{-\infty}^{\tau_n} d\tau_{n-1} \times \cdots 
\nonumber\\
\nn\\
&\ & \hspace{-15mm} \times \cdots \times\  \int_{-\infty}^{\tau_2} d\tau_1\ 
\exp {\left[\;  -\ \frac{i}{\hbar}\ \sum_{{\rm states}\, b=0}^{n-1}
\left(\, {\sum_{j\, {\rm in}\, b} E(\vec p_j)}\right) (\tau_{b+1}-\tau_{b})\right]}\, ,
\label{eq:topt-2}
\eea
where the $a$th factor $V_{a\leftarrow a-1}$ is the matrix element of Eq.\ (\ref{eq:potential-element}) that takes state $|m_{a-1}\rangle$ into the state $|m_a\rangle$, with $|m_n\rangle \equiv |m_{\rm out}\rangle$.  In fact, this expression is given in the interaction picture, where we remove the phases associated with $|m_0(-\infty)\rangle$ and $|m_{\rm out}(\infty)\rangle$, thus dropping two of the terms in the phase proportional to $\tau_0$ and $\tau_{n+1}$. The sums over states are suitably-normalized integrals over free-particle phase space, which we may reinterpret as integrals over the loop momenta of a sum of perturbation theory (Feynman) diagrams.
  
 Generically, the sums over states in Eq.\ (\ref{eq:topt-2}) are divergent whenever subsets of $\tau$'s coincide, $\tau_i\rightarrow \tau_j$, and also when some set of $\tau_i$ go to infinity.   The former, ``UV" problem is handled by renormalization, and the solution is summarized by scaling each term in $V$ by an appropriate coupling constant $g(\mu)$, with $(\tau_i-\tau_j)_{\rm min}=1/\mu$.   In 4 dimensions, only Yang-Mills theories have the property of asymptotic freedom, $g(\mu) \sim 1/\ln(\mu)$ \cite{Gross:1973id,Politzer:1973fx}.  The couplings of the Standard Model are either asymptotically free, or are small enough to not change much over experimentally-accessible energies.  This makes an expansion in powers of $\alpha(\mu)=g^2(\mu)/4\pi$ plausible, at least in principle. 
 
 Once we do the expansion to calculate the amplitude for a process, the form of an ``ideal cross section", the square of the amplitude, would  be one with only a single kinematic scale, to which we can set  renormalization scale $\mu$,
\bea
 Q^2\; \hat \sigma_{\rm SD}(Q^2,\mu^2,\alpha_s(\mu)) 
&=&
\sum_n c_n(Q^2/\mu^2)\; \as^n(\mu) + {\cal O}\left({1\over Q^p}\right)
\nonumber\\
&=& \sum_n c_n(1)\; \as^n(Q) +   {\cal O}\left({1\over Q^p}\right)\, ,
\label{eq:irs}
\eea
up to corrections that vanish as $Q^{-p}$, for some positive power $p$,  typically an integer. 
The key is to find quantities that are observable, and for which the coefficients are well-behaved, and do not depend on scales for which the coupling is too large.  Such quantities are sometimes called ``infrared safe".   For proton accelerators or hadronic final states, the problem is that there are essentially no cross sections that  qualify as infrared safe without further analysis.   What is reason for this problem?

\subsection{Mass-shell enhancements in perturbation theory} 

As we have seen, the Schr\"odinger equation gives transition matrix elements as sums of  ordered time integrals.
These time integrals extend to infinity, but usually oscillations damp them and they provide finite answers.  Long-time, ``infrared" divergences (logs) come about only when phases vanish so that the time integrals diverge.

When does this happen?  We can tell by reorganizing the phase in Eq.\ (\ref{eq:topt-2}),
\bea
\exp \left[\;  -\ \frac{i}{\hbar}\ \sum_{{\rm states}\, b}
\left(\, \sum_{j\, {\rm in}\, b} E(\vec p_j)\right) (\tau_{b+1}-\tau_b)\right]
&=&
\nonumber\\
&\ & \hspace{-55mm} 
\exp \left[\;  -\ \frac{i}{\hbar}\ \sum_b
\left(\sum_{j\, {\rm in}\, b-1} E(\vec p_j)\ -\ \sum_{j\, {\rm in}\, b} E(\vec p_j)\right)\ \tau_b\, \right]\, .
\label{eq:phases}
\eea
Divergences can occur in the integrals of Eq.\ (\ref{eq:topt-2}) for $\tau_i\rightarrow\infty$ if two requirements are met.
\begin{enumerate}
\item As is shown by the right-hand side of (\ref{eq:phases}), the phase must vanish, corresponding to sequences of degenerate states, 
\bea
\sum_{j\, {\rm in}\ b-1} E(\vec p_j)\ =\ \sum_{j\, {\rm in}\ b} E(\vec p_j)\, .
\label{eq:degeneracy}
\eea
\item From the left-hand side, even if it vanishes, the phase must also be stationary with respect to the momentum integrals that are implicit in the sums over states in Eq.\ (\ref{eq:topt-2}).  
 Imposing momentum conservation, this translates into the vanishing of their derivatives with respect to loop momenta $\ell_i^\mu$,
\bea
{\partial \over \partial \ell_i{}_\mu} \; \left[\, {\rm phase}\, \right] 
=
 \sum_{{\rm states}\, b}\ \sum_{j\, {\rm in}\, b} \, (\eta_{ij} \beta_j^\mu)(\tau_{b+1}-\tau_b)\ =\ 0\, ,
\label{eq:stationary}
\eea
where the $\beta_j$s are 4-velocities, 
\bea
\frac{\partial E({\vec p}_j)}{\partial \ell_{i\mu}}\  =\ \frac{\partial p_{j\mu}}{\partial \ell_{i\mu}}\; \frac{\partial E({\vec p}_j)}{\partial p_{j\mu}} \ =\  \eta_{ji}\;  \beta_j^\mu\, ,
\label{eq:velocity}
\eea
with $\eta_{ji}=\pm 1,0$, depending on whether loop $i$ flows on line $j$, and if so whether  momentum $\ell_i$ flows in the same or opposite direction as $p_j$. 

\end{enumerate}

Now, any vector of the form $\beta_j^\mu \Delta \tau = \Delta x^\mu$ is a translation that is consistent with free-particle classical equations of motion.  Equation (\ref{eq:stationary}) implies that around each loop, every sequence of free propagations is consistent.   Infrared divergences, then, arise from regions in the sums over states where particles describe free, classical propagation that extends to $t\rightarrow \infty$, even as the numbers of particles may change through the actions of vertices.  This is easy to satisfy in subdiagrams where all the $\beta_j$'s are equal, but is otherwise quite restrictive.  Thus, whenever fast partons emerge from the same point in space-time, amplitudes are enhanced by corrections that describe the rescattering, creation and absorption of collinear partons at large times.
 
\begin{figure}
\centerline{\figscale{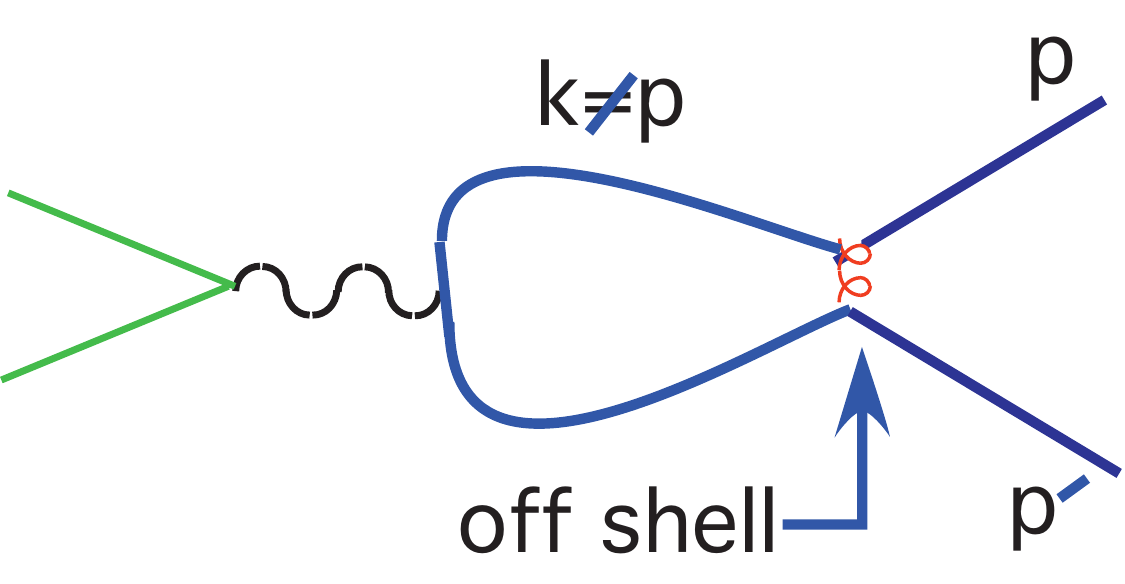}{5cm}}
\caption{Example of a degenerate intermediate state that cannot give long-time divergences. \label{fig:no-pinch}}
\end{figure}

\begin{figure}
\centerline{\figscale{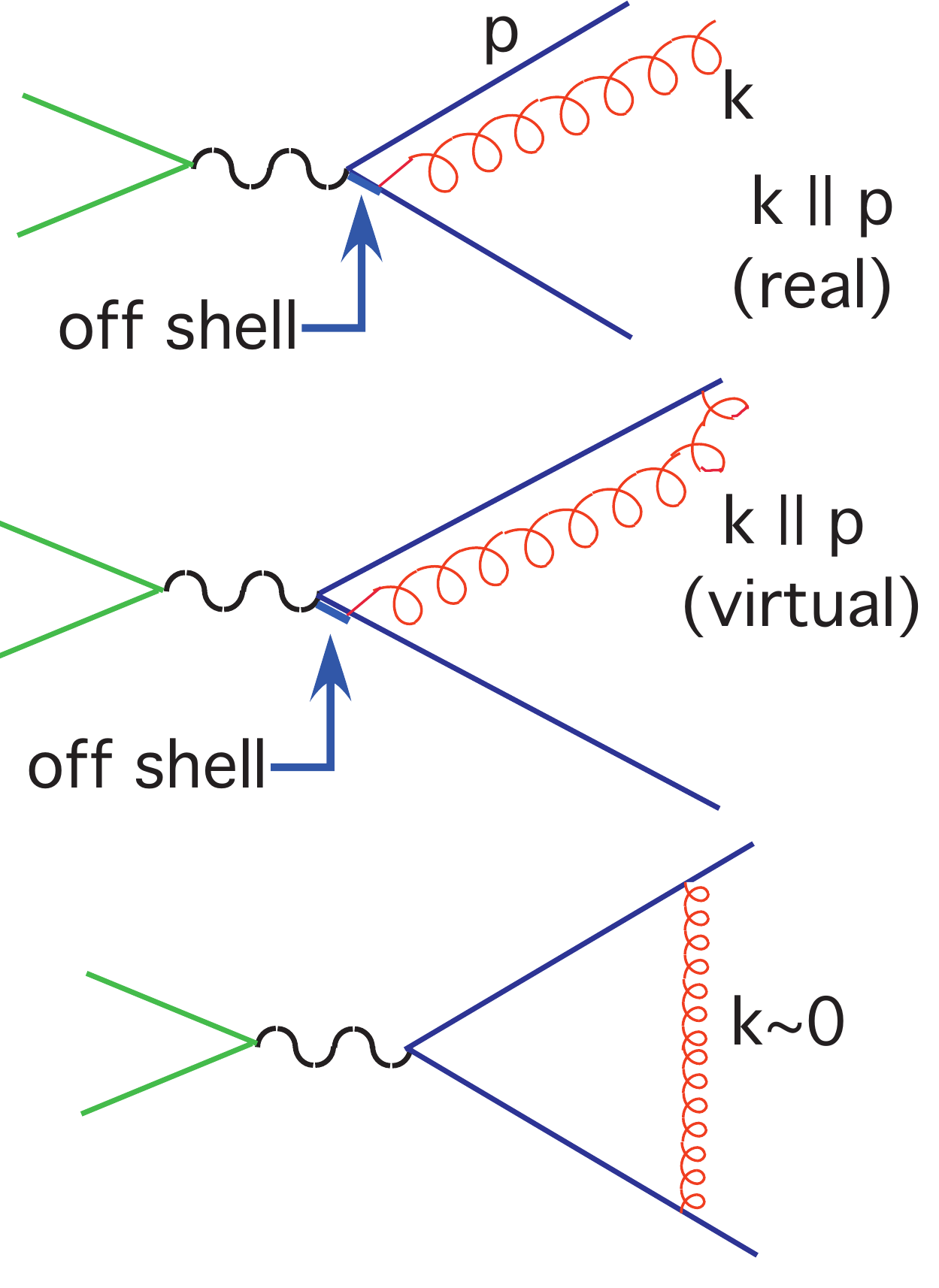}{4cm}}
\caption{Low order pinch surfaces for electroweak boson decay. \label{fig:epem}}
\end{figure}

A simple example, in Fig.\ \ref{fig:no-pinch}, illustrates the surprising power of the requirement of classical propagation.   The intermediate state involving momentum $k$ may be degenerate with the final state, but if the two lines emerging from the decay are not parallel, they can never meet again though free-particle propagation.   Hence these degenerate states are not associated with a stationary point in the integral, and there is no possibility of an infrared divergence.
This kind of reasoning makes identifying infrared enhancements a lot simpler.   For particles emerging from a local scattering, (only) collinear or soft (infinite wavelength) lines can give long-time behavior and enhancement. The most straightforward  examples of configurations that do give divergences are in the hadronic decay of electroweak bosons, or $\rm e^+e^-$ annihilation in the single-electroweak boson approximation, illustrated in Fig.\ \ref{fig:epem}.
 This pattern generalizes to any order, and any field theory, but gauge theories are the only renormalizable theories with soft ($k\rightarrow 0$) divergences.

 For $\rm e^+e^-$ annihilation, this implies by the optical theorem that the total cross section is infrared safe \cite{Appelquist:1973uz}. To lowest order in electroweak interactions, but to all orders in QCD, the total cross section is proportional to the imaginary part of the electroweak boson self energy.   The same reasoning that applies to Fig.\ \ref{fig:no-pinch} shows that there are no stationary points with finite-energy on-shell lines at any loop order.   Because the self energy is finite, so is its imaginary part, and thus so is the total cross section.   The same reasoning, based on the optical theorem, also applies to  jet cross sections \cite{Sterman:1977wj,Sterman:1978bi}, all of whose singularities can be derived from a rotationally non-invariant, but still hermitian, truncation of the quantum field theory Lagrangian.
 
 \subsection{Jets:  Rare but Highly Structured Events.}

\begin{figure}
\centerline{  \figscale{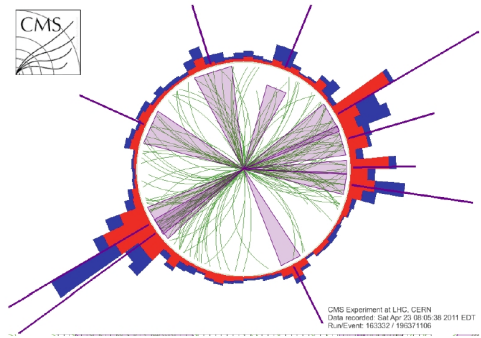}{7 cm}  }
\caption{A multi-jet event as observed by the CMS experiment at the Large Hadron Collider. \label{fig:cms-jets}}
\end{figure}

 At energies much above the mass of the nucleon, certain events include subsets of particles $\{q_j\}$ with anomalously small  invariant mass, $( \sum_i q_i )^2 \ll ( \sum_i E_i )^2$, and  not embedded among other particles of similar energy.   Such sets are  ``jets", and, as depicted in Fig.\ \ref{fig:cms-jets} from the CMS experiment at the Large Hadron Collider, a single scattering event may produce a number of jets.
Events containing jets, in which the flow of momentum is radically changed between initial and final states, are a signature of large momentum transfers through local interactions, and as such direct evidence of processes taking place on distances of the order of 1/(momentum transfer).      

The history of the term ``jets" applied to final states in particle collisions goes back to  the 1950's, with sprays of particles observed in collisions of cosmic rays with detector materials.   In Ref.\ \cite{Edwards:1957}, we read ``The average transverse momentum resulting from our measurements is $p_T$=0.5 BeV/c for pions \dots [a table] gives a
 summary of jet events observed to date \dots".   These jets emerged from cosmic rays with  energies far above what could then be achieved in the laboratory.  The jets of these events are by now interpreted as fragments of the projectile nucleus, whose collider analogs are sometimes referred to as ``beam jets".   They are not, for the most part, a signal of very  large momentum transfer processes, or of the decay of newly-created heavy particles.   The observation of jets of the latter types had to wait for the era of high energy physics and the discovery of the Standard Model.
  
This modern story begins with the parton model for inclusive deep inelastic electron-nucleon scattering (DIS) \cite{Bjorken:1969ja}.   In the parton model, this {\it inclusive} process is approximated by the {\it exclusive} scattering of an electron by a charged constituent of the nucleon, multiplied by a function $F(x)$ that depends only on a ``scaling variable", $x\equiv {Q^2}/{2p\cdot q}$, with $q$ the 4-momentum transfer and $Q^2\equiv -q^2>0$,
\bea
\sigma_{e\, {\rm proton}}^{\rm incl}\left(Q,x=\frac{Q^2}{2p\cdot q}\right) \ \rightarrow\  \sigma^{\rm excl}_{e\, \rm parton} (Q,xp) \times F_{\rm proton}(x)\, .
\label{eq:scaling}
\eea
Variable $x$ is interpreted as the fractional momentum of the nucleon carried by the parton, when all masses are neglected.   The elastic electron-parton scattering is calculated to lowest order in quantum electrodynamics,
\bea
e(k)\ +\ a(xp)\ \to\ e(k-q)\ +\ a(xp+q)\, ,
\label{eq:e-parton}
\eea
where $a$ represents the parton, a quark or antiquark in QCD.   The value of $x$ is determined simply by requiring the scattered parton to remain massless, $(xp+q)^2=0$.  In the parton model, the function $F(x)$ has the interpretation of a probability distribution of momentum fractions for parton $a$ in the nucleon.  Its independence of the momentum transfer is known as ``scaling".   Scaling turned out to be a striking and successful description of early DIS data.   Its explanation in terms of asymptotic freedom \cite{Gross:1973id,Politzer:1973fx} was an electrifying development in the discovery of the Standard Model.   For an asymptotically free theory, the coupling becomes weak at short distances, even as it increases at longer distances.   This made sense for the inclusive cross section, but the question arises as to what happens to partons in the final state?  Doesn't confinement forbid a direct phenomenological expression for quarks?  

The unequivocal answer, ``no", came from SLAC in 1975:   In electron-positron annihilation to hadrons, the angular distribution for energy flow follows the Born expression for the creation of spin-1/2 pairs of fermions: the quarks and antiquarks \cite{Hanson:1975fe}.    Jets are ``rare" because the high momentum transfer scattering of partons is rare, but once a hard scattering has occurred they are inevitable, and the rates of their appearances are calculable.

\begin{figure}
\centerline{\figscale{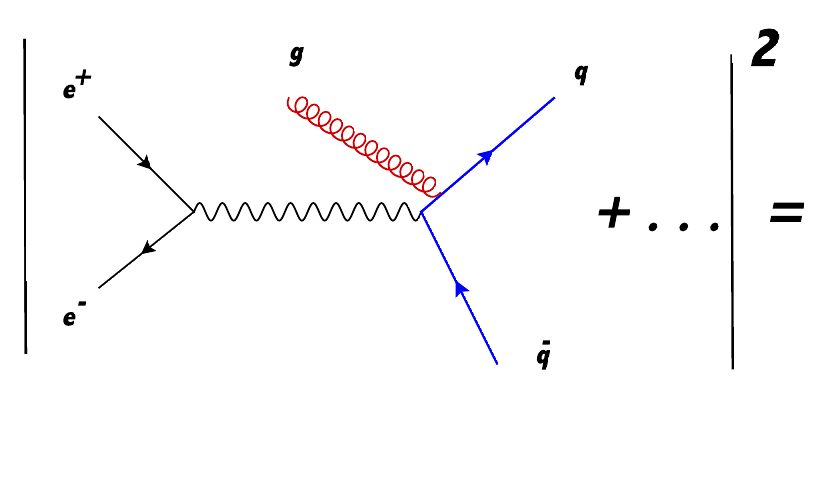}{6.4 cm} \quad
\figscale{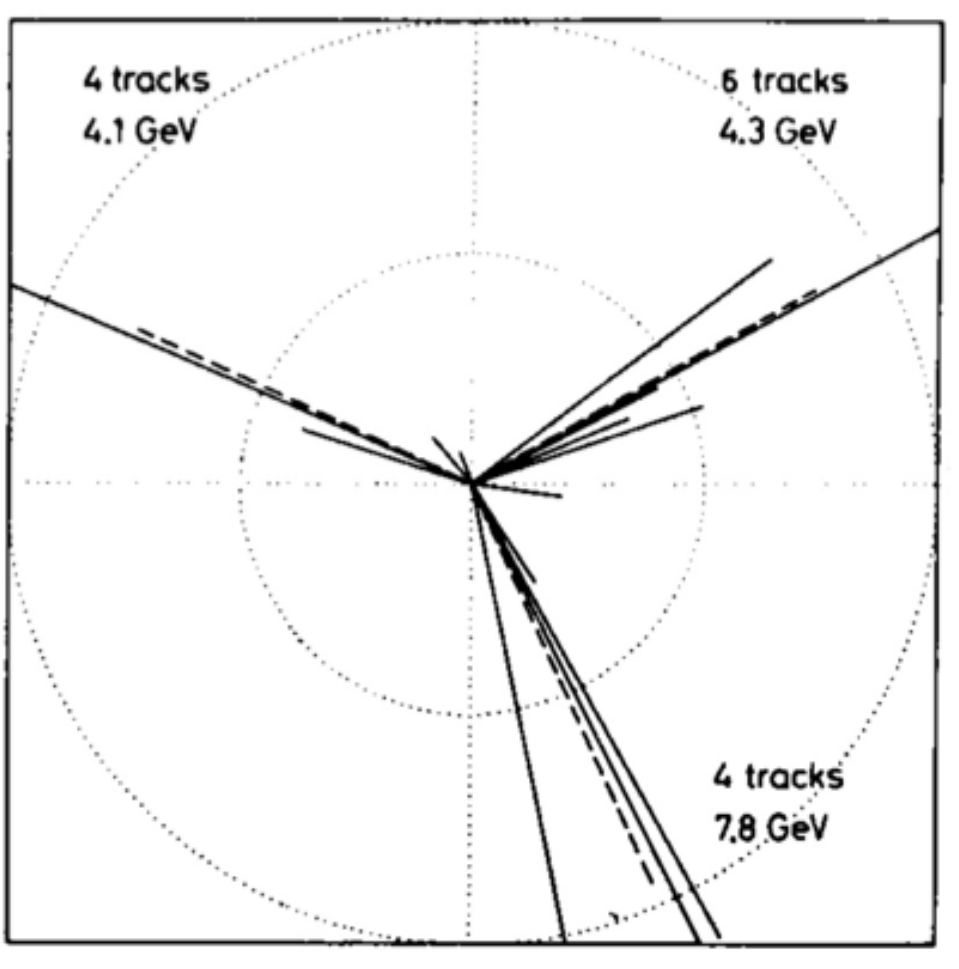}{4.8 cm} }
\caption{The partonic process for a three jet event in lowest order QCD and the corresponding TASSO experiment signal at the Petra accelerator \cite{Wu:1984ik}.}
\label{fig:3-jet}
\end{figure}

 After the quark jets of SLAC, came hints of  jets associated with gluons in Upsilon decay \cite{Berger:1978rr},  and by 1979 clear gluon jets at Petra 
\cite{Wu:1984ik,Ellis:1976uc,Ellis:1978wp}, illustrated in Fig.\ \ref{fig:3-jet}.

To compute the probabilities of such events we compute jet cross sections directly in perturbative QCD as though the final state consistsed of quarks and gluons \cite{Sterman:1977wj,Sterman:1978bi}  Such a prescription, purely in terms of partonic language, seemed strange at first, knowing that quarks and gluons are confined.   Nevertheless, the theory gives a prediction, and the theory can tell us when this prediction is not self-consistent.  That is, we assume that infrared safe perturbation theory provides an asymptotic expansion.

\begin{figure}
\centerline{ \figscale{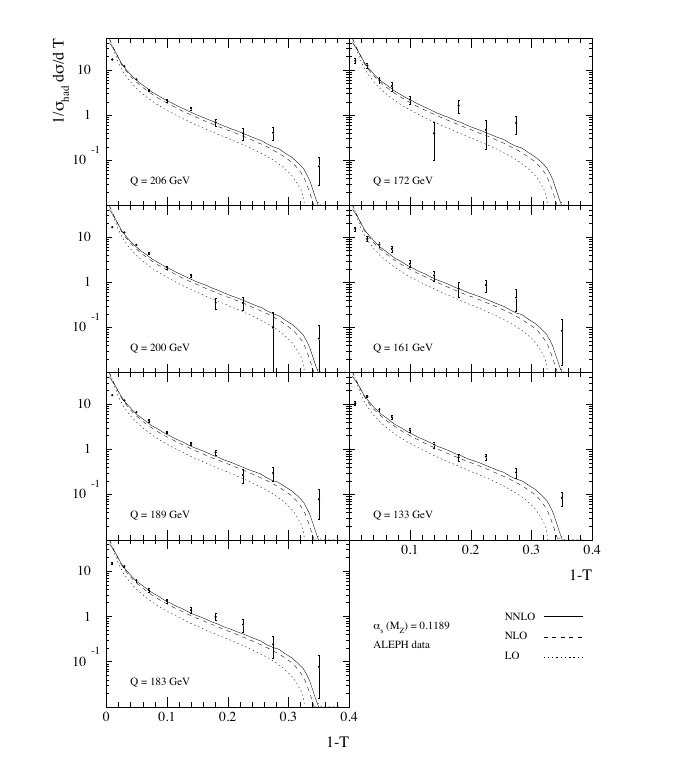}{9cm}}
\caption{Data vs.\ theory for the thrust \cite{GehrmannDeRidder:2007hr}.}
\label{fig:thrust-data}
\end{figure}

A good illustration of the kind of cross section we can calculate  is the ``thrust"\cite{Farhi:1977sg}, in $\rm e^+e^-$ annihilation, defined by 
\bea
T &\equiv& \max_{\hat n}\  \frac{1}{Q}\ \sum_{{\rm particles}\ i}\ \vec p_i\cdot \hat n\, ,
\label{eq:thrust}
\eea
where the maximum is taken over all unit directions $\hat n$ in the sphere of the center-of-mass frame.   The thrust equals unity for perfectly pencil-like, back-to-back jets of massless particles.   By now, the coefficients of $\alpha_s/\pi$ (lowest order, or LO), $(\alpha_s/\pi)^2$ (next-to-lowest, NLO) and $(\alpha_s/\pi)^3$ (next-to-next-to-lowest, NNLO) are known \cite{GehrmannDeRidder:2007hr}, and agreement with experiment extends over several orders of magnitude, as illustrated in Fig.\ \ref{fig:thrust-data}.

\subsection{Factorization}

Machines for hadron-hadron scattering allowed for the scattering of quarks and gluons from pre-existing hadrons, whose internal interactions long predate their entrance into a collider beam -- indeed, they generally extend back to baryosynthesis in the early universe!   Such interactions, involving confinement, are clearly not calculable in perturbation theory.   In this case, we apply the method of factorization, in which perturbatively calculable short-distance effects are separated from long-distance nonperturbative dynamics, which enter as parton distributions.    Parton distributions are the pre-existing probability densities for individual quarks and gluons to carry the fractional momenta of the colliding nucleons.   Factorization is the key to predictions for proposed and established  theories, and in general takes the form
\bea
\hspace{-12mm}
Q^2\sigma_{\rm phys}(Q,m)
&=&
\hat\sigma (Q/\mu,\as(\mu))\, \otimes\, f_{\rm LD}(\mu,m) 
\ + \ {\cal O}\left(\frac{1}{ Q^p}\right)\, ,
\label{eq:basic-fact}
\eea
which can be thought of as a generalization of both Eqs.\ (\ref{eq:irs}) and (\ref{eq:scaling}).   
In expressions like this, $\mu$ is referred to as the factorization scale, which is the lowest momentum scale on which the ``short-distance" function $\hat \sigma$ depends, and $m$ represents the infrared scales whose dependence is factorized into the ``long-distance" function $f_{\rm LD}$.    ``New physics" is in  $\hat\sigma$, while $f_{\rm LD}$  is ``universal".  Factorizaton of this sort is required for almost all collider applications, and the determination of parton distributions requires a synthesis of measurements of many factorized cross sections \cite{Rojo:2015acz}.  The lowest-order process of (\ref{eq:scaling}) is now generalized beyond electron-quark scattering, and dressed in $\hat\sigma$ by QCD quantum corrections, as in (\ref{eq:irs}).   In place of the simple product in (\ref{eq:scaling}), the product
$\otimes$ represents a convolution in terms of partonic fractional momenta, or other kinematic degrees of freedom that are not observed directly.

What we actually do is to compute the ``physical cross section" $\sigma$ on the left-hand side of (\ref{eq:basic-fact}) and $f_{\rm LD}$ on the right-hand side, in an infrared-regulated variant of QCD, where we can prove the factorization explicitly.    We then extract the perturbative quantity $\hat \sigma$, assuming it is the same for true QCD as for its IR-regulated cousin.  Factorization for a given cross section generally requires that it be sufficiently inclusive that no small parameters are introduced in the selection of final states.   The form of factorization may also depend on the measurement.   Calculations of the short distance functions generally become very complex beyond the lower order in $\alpha_s$.  An enormous amount of work has been put into the calculation of the perturbative amplitudes \cite{Berger:2009zb,Elvang:2013cua} upon which these cross sections are based, which has led to fruitful interplay between QCD phenomenology, abstract quantum field theory and pure mathematics.  The steps from amplitudes to cross sections with prescribed phase space presents further challenges.   Although great progress has been made over the past few years, much remains to be done to exploit the full potential of accelerator data \cite{Campbell:2013qaa}.  A recent milestone is the completion of the order $\alpha_s^3$ corrections to $\hat \sigma$ for the inclusive production of the Higgs boson in the Standard Model \cite{Anastasiou:2013mca}.     

\section{Evolution/Resummation}

The full power of asymptotic freedom and factorization requires evolution, by which we can control dependence on the factorization scale $\mu$ in factorized cross sections like (\ref{eq:basic-fact}).   This enables us to compute the short distance scattering $\hat\sigma$ in terms of a single scale, with $\mu\sim Q$, and hence to derive the most accurate expansions in $\alpha_s(Q)$ as $Q$ increases. Whenever there is a factorized physical quantity, we can derive an evolution equation, starting with the independence of any such  quantity from the factorization scale,
\bea
0=\mu{d\over d\mu} \ln \sigma_{\rm phys}(Q,m)\, .
\label{eq:mu-indep}
\eea
For simplicity, suppose $\sigma_{\rm phys}$ is a simple product of $\hat\sigma$ and $f$.   Then by an elementary separation of variables we find that
\bea
\mu{d \ln f\over d\mu}= - P(\as(\mu)) = - \mu{d \ln \hat\sigma \over d\mu}\, ,
\label{eq:separate}
\eea
where the function $P(\as)$ depends on only the (dimensionless) variables held in common by the short- and long-distance functions.  We can calculate $P(\as)$, which may be referred to as a splitting function or an anomalous dimension depending on the context, because it is the derivative of $\hat\sigma$, which we can calculate.    Equations of this sort are said to describe evolution or resummation.  Applied to Eq.\ (\ref{eq:basic-fact}), for example, Eq.\ (\ref{eq:separate}) implies that
\bea
 \sigma_{\rm phys}(Q,m) = \sigma_{\rm phys}(q,m)\ \left[ \frac{\hat \sigma(Q,\alpha_s(Q)}{\hat\sigma(q,\alpha_s(q))} \right]\ \exp\left\{  \int_q^Q {d\mu'\over \mu'} \, 
P\left( \alpha_s(\mu')\right) \right\}\, ,
\label{eq:sigma-evolve}
\eea
in which the second and third factors on the right-hand side, which contain all $Q$-dependence, are computable in perturbation theory so long as $\alpha_s(Q)$ and $\alpha_s(q)$ are both small.   This means that in a theory with asymptotic freedom, observations at moderate scales, $q$ say, lead to predictions for all larger scales, $Q$.

 Such factorization, and hence evolution applies, for example, to the dimensionless structure function $F_2(Q^2,x)$ in DIS, which can be thought of as a cross section with dimensionfull kinematic factors removed.   Specifically, we consider its Mellin moments with respect to the scaling variable, $x$, $\tilde F_2(Q^2,N)=\int_0^1 dx\ x^{N-1} F_2(Q^2,x)$.   These moments resolve the convolution in partonic momentum fraction into a product of the form
 \bea
 \tilde F_2(Q^2,N)\ =\ \sum_{{\rm partons}\ a} C_{2a}\left( N,\frac{Q}{\mu},\alpha_s(\mu)\right)\ f_a(N,\mu)\, .
 \eea
Here, the sum is over parton flavors:  quarks, antiquark, gluons.   On the right, the $C_{2a}$ are  short-distance ``coefficient functions", one for each parton flavor, and the $f_a(N,\mu)$ are Mellin moments of parton distributions, $f_a(\xi,\mu)$, in the spirit of Eq.\ (\ref{eq:basic-fact}), with respect to momentum fraction, $\xi$.  In this case, the separation constants $P_{ab}(N,\as)$ depend on the moment variable $N$ and are denoted $\gamma_{ab}(N,\as)$.
This is a matrix, as variations with the factorization scale reflect the effects of quark pair creation and gluon radiation, which can change the flavor of the parton that initiates the hard scattering.   The evolution equation for the long-distance parton distributions can then be written as
\bea
\mu\frac{\partial}{\partial \mu}\ f_a(N,\mu)\ =\  \sum_b\ \gamma_{ab}(N,\alpha_s(\mu))\ f_b(N,\alpha_s(\mu))\, ,
\eea
which is one of the forms of the celebrated Dokshitzer-Gribov-Lipatov-Altarelli-Parisi (DGLAP) equation \cite{Gribov:1972ri,Altarelli:1977zs,Dokshitzer:1977sg}, the bedrock of high energy phenomenology at colliders.

\begin{figure}
\centerline{ \figscale{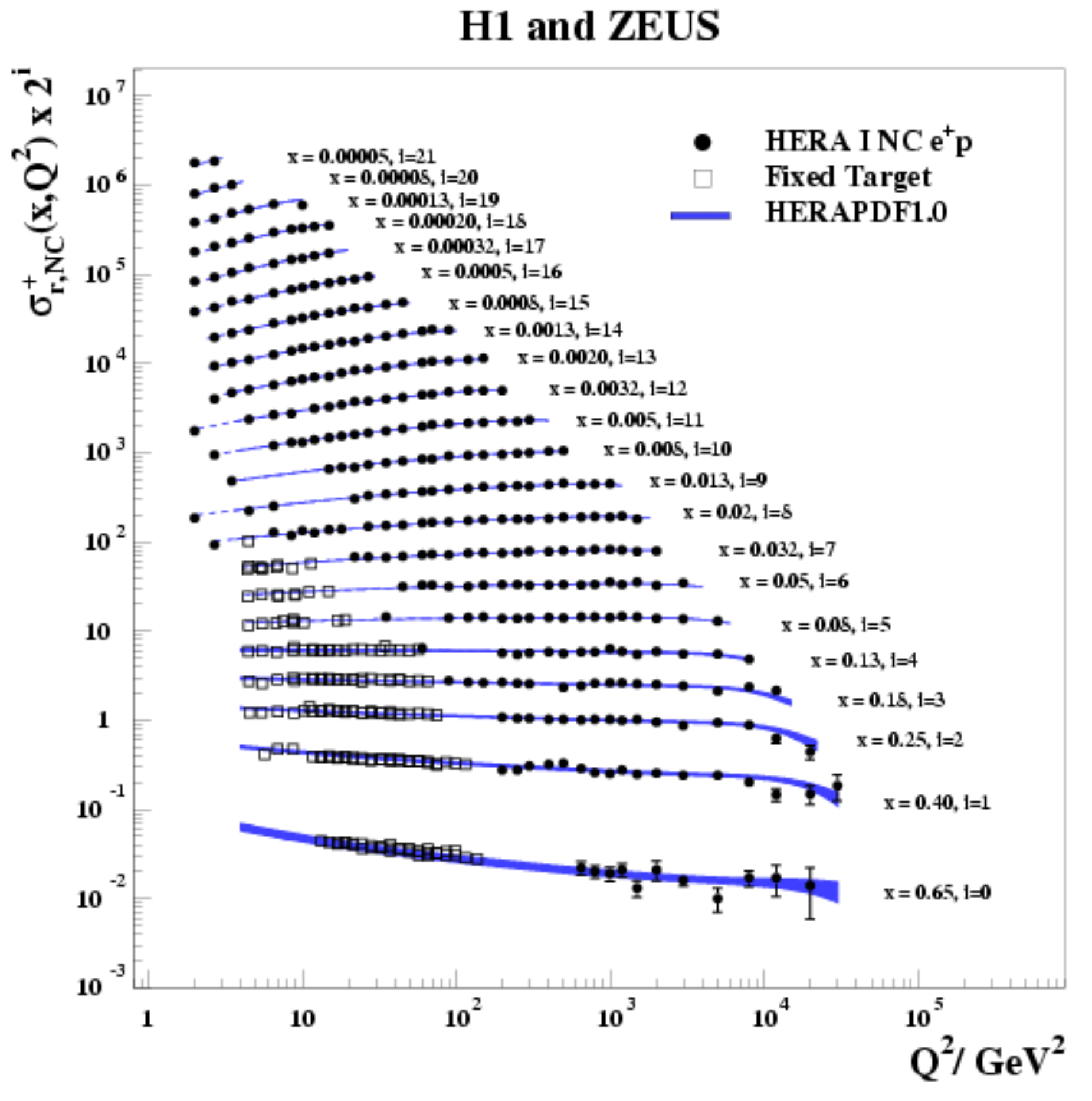}{8 cm} }
\caption{DIS cross sections from Zeus and H1 experiments at the HERA accelerator \cite{Aaron:2009aa}.}
\label{fig:evol}
\end{figure}

If for simplicity we suppress the matrix structure, and neglect the analog of the perturbative prefactor in Eq.\ (\ref{eq:sigma-evolve}), we find for our structure functions the $Q$-dependence,
\begin{eqnarray}
\hspace{-12mm}
{\tilde F_2(Q^2,N)}
= {\tilde F_2(Q^2_0,N)}
\; \exp\left[\;  -{1\over 2}\; \int_{Q_0^2}^{Q^2}\;
{d\mu'{}^2\over \mu'{}^2}\ \gamma(N,\alpha_s(\mu'))\; \right]\, ,
\eea
again enabling the use of lower-energy observations to give higher energy predictions.
Expanding $\gamma(N,\as)=\gamma_N^{(1)}(\as/\pi)+ \dots$, and using the lowest-order version of the running coupling, $\as(\mu) = 4\pi/b_0\ln(\mu^2/\Lambda_{\rm QCD}^2)$, we find the approximation
\bea
\hspace{-12mm}
{\tilde F_2(Q^2,N)}\
= {\tilde F_2(Q_0^2,N)}
\left({\ln(Q^2/\Lambda_{\rm QCD}^2)
\over 
\ln(Q_0^2/\Lambda_{\rm QCD}^2)} \right)^{-2\gamma_N^{(1)}/b_0 }\, .
\end{eqnarray}
In its full form, which depends on all of the particle content and dynamics of quantum chromodynamics, this procedure works really well.  It implies the approximate scaling seen in early DIS experiments at moderate $x$ and pronounced evolution for smaller $x$, as shown in Fig.\ \ref{fig:evol}.

Hard hadron-hadron scattering, as at the LHC, is mediated by partons from both colliding hadrons, and requires only a modest extension of the factorization formalism.   For  an inclusive process in hadron-hadron scattering that involves a large momentum transfer $M$ to produce final state $F(M)+X$, the corresponding factorized cross section can be written as
\bea
d\sigma_{\rm H_1H_2}(p_1,p_2,M) &=&\ 
\sum_{a,b} \int_0^1 dx_a\, dx_b\
d\hat{\sigma}_{ab\to F+X}\left(x_a p_1,x_bp_2,M,\mu\right)\;
\nn\\
&\ & \hspace{30mm} \times
 f_{a/H_1}(x_a,\mu)\, f_{b/H_2}(x_b,\mu) ,
\label{eq:hh-fact}
\eea
in terms of a perturbative short distance (differential) cross section $d\hat \sigma$, combined with the same parton distribution functions as above, but evolved to higher scales in general.  This form is a straightforward variant of the generic factorization (\ref{eq:basic-fact}), but of course, it requires a proof.   Factorization proofs, which have been the subject of  considerable effort,  justify the ``universality" of the parton distributions, that is, that parton distributions measured in deep-inelastic scattering are the same functions that appear in hadron-hadron scattering.  Because this is indeed the case, data from one class of experiments can be used to make predictions for another class.   Thus, for example, the measurements of parton distributions at HERA in the 1990s can be used to predict cross sections for hypothetical super-partners at the LHC in Run 2.  

We will come back to the how and why of factorization, which is fundamental to applications of gauge theories at accelerators, in the closing pages.   We can use jet cross sections to illustrate the success of the factorization paradigm.   

\begin{figure}
\centerline{\figscale{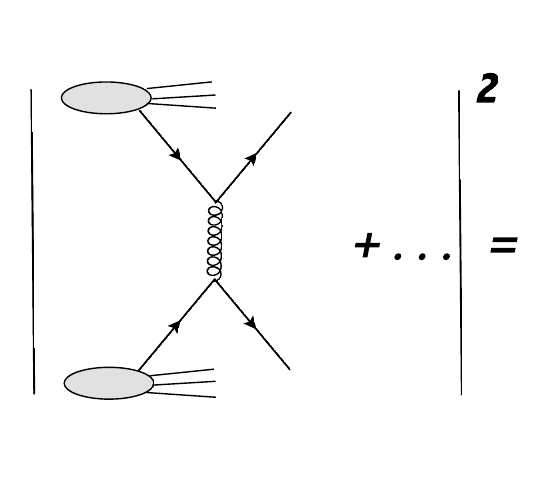}{5cm}\ \ \ \figscale{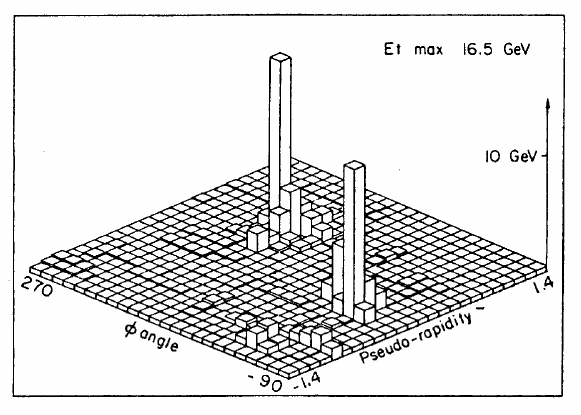}{7cm}}
\caption{Equivalence of QCD quark-quark scattering and the observation of jets by the flow of energy in the UA1 detector\cite{Arnison:1983rn}.}
\label{fig:sps-lego}
\end{figure}

  Through the early 1980's,  there were strong suggestions of scattered parton jets at CERN and Fermilab \cite{Della Negra:1977sk,Corcoran:1979xt}.  Fuller clarification, however, awaited experiments at the SPS collider, whose large angular coverage made possible plots that exhibited energy flow over the whole detector.   This led to the observation of high-$p_T$ jet pairs that unequivocally represent the scattering of partons at short distances.   The underlying equivalence of the quantum mechanical picture of gluon exchange between quarks and detector signals is represented by Fig.\ \ref{fig:sps-lego}, in which the towers on the ``unrolled" detector surface reveal the flow of energy \cite{Banner:1982kt,Arnison:1983rn}.

\begin{figure}
\centerline{\figscale{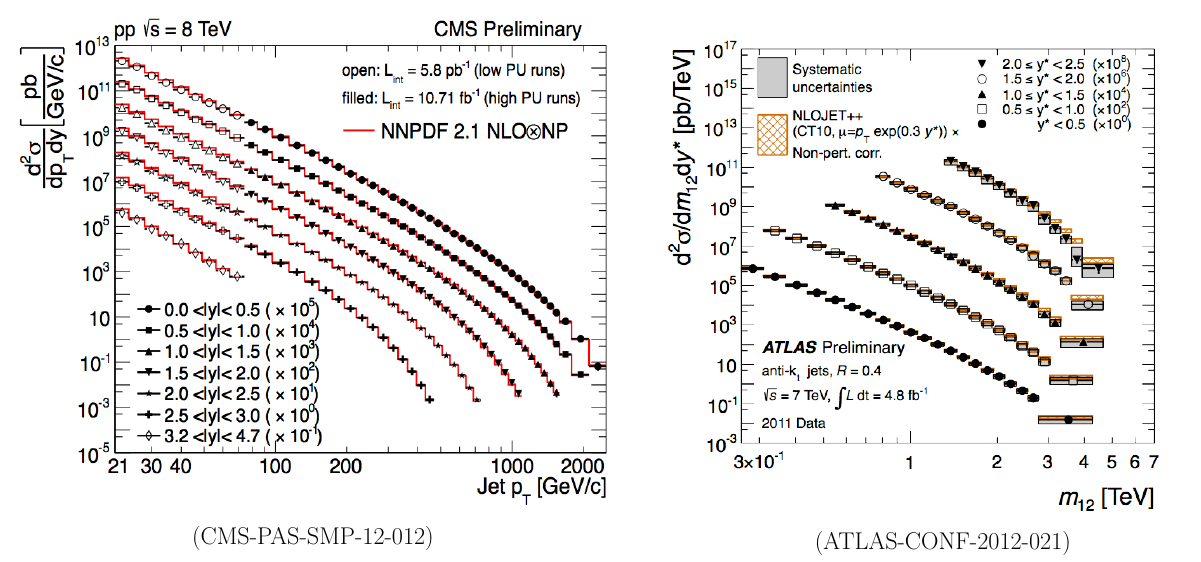}{12cm}}
\caption{Jet $p_T$ distributions measured by the CMS experiment, and pair mass distributions measured by the ATLAS experiment, showing agreement with Standard Model predictions over many orders of magnitude.}
\label{fig:jetcrossec}
\end{figure}

 The nineteen nineties ushered in what we may call the era of the great ``Standard Model Machines".   HERA for DIS, LEP I and II  for $\rm e^+e^-$ annihilation, and the Tevatron Run I for $p\bar p$ collisions together provided jet cross sections over multiple orders of magnitude, as they established and confirmed the electroweak structure and flavor content of the Standard Model.   The current decade has brought a new era of jets at the limits of the Standard Model, initiated by Tevatron Run II, and realized with the ongoing Large Hadron Collider program, at 7, then 8, and now 13 TeV in the center of mass.  Events transpiring at the scale $\delta x \sim \frac{\hbar c}{1 \ {\rm TeV}} \sim 2\times 10^{-19}$ meters are now routinely observed about 10 meters away, an observational bridge of twenty orders of magnitude.   The impressive success of theory predictions based on factorized cross sections like Eq.\ (\ref{eq:hh-fact}) is shown in Fig.\ \ref{fig:jetcrossec}.

At this point, we may mention another life for jets,  ``shining from the inside" as a probe for new phases of strongly-interacting matter in nuclear collisions  \cite{Adcox:2004mh,Adams:2005dq} at the Relativistic Heavy Ion Collider (RHIC) and the LHC, and, prospectively in ``cold nuclei" at a electron-ion collider \cite{Accardi:2012qut}.   

With this motivation, we turn to the physics behind, and the arguments for, the factorization properties on which the calculation of jet and related cross sections are based.

\section{Gauge Theory Factorization}

Understanding factorization is about learning how to calculate with a theory that acts differently on different scales.  For the purposes of  jet cross sections, the underlying factorization takes a form first recognized in the late 1970s \cite{Libby:1978qf,Ellis:1978ty,Efremov:1978xm,Sachrajda:1978ja,Mueller:1978xu,Mueller:1981sg}.
\bea
{d\sigma(A+B\rightarrow \{p_i\}) }\
&=&  \sum_{a,b}\ \int_0^1  dx_adx_b \ {f}_{a/A}(x_a,\mu_F)\, {f}_{b/B}(x_b,\mu_F)\ 
 \nonumber\\
 \ \nn \\
 &\ & \hspace{5mm}
 \times \ \ dC\left(x_ap_A,x_bp_B,\frac{Q}{\mu_F},\frac{p_i\cdot p_j}{p_k\cdot p_l}\right)_{ab\rightarrow c_1 \dots c_{N_{\rm jets}+X}}
 \nonumber\\
 &\ & \nn \\
 &\ & \hspace{5mm}
  \times \ \left[ \prod_{i=1}^{N_{\rm jets}}\ J_{c_i}(p_i,\mu_F) \right]\, .
  \label{eq:jet-fact-hg}
\eea
 Like Eqs.\ (\ref{eq:basic-fact}) and (\ref{eq:hh-fact}), this formula includes parton distributions and short-distance ``coefficient" functions, as well as a further factorization into  functions that represent the jets.   Closely-related factorizations apply to processes involving exclusive final states \cite{Efremov:1979qk,Lepage:1980fj,Ji:1996nm}, and have found renewed development in the language of effective field theory \cite{Bauer:2000yr,Bauer:2001yt,Becher:2014oda}.
 
 Expressions of this form tell a story of nearly on-shell propagations in the initial and final states, punctuated by a single short-distance interaction.   As mentioned above, the definitions of the jets must be sufficiently inclusive that no small scales are introduced.

The jets themselves are characterized by correlated internal dynamics, which is  ``autonomous" relative to the remainder of the process.   We have already seen that enhancement of correlations between collinear particles is built into quantum field theory.  Where does the automonomy come from?

 To distill the essence of this argument, we will think of classical fields seen by scattered charges \cite{Basu:1984ba}.  Even though we are working in quantum field theory, a classical picture isn't so far-fetched, because the correspondence principle is the key to infrared divergences in perturbation theory.   An accelerated charge must produce classical radiation, and an infinite number of soft gauge vector bosons is required to  make a classical field.   

\begin{figure}
\centerline{ \figscale{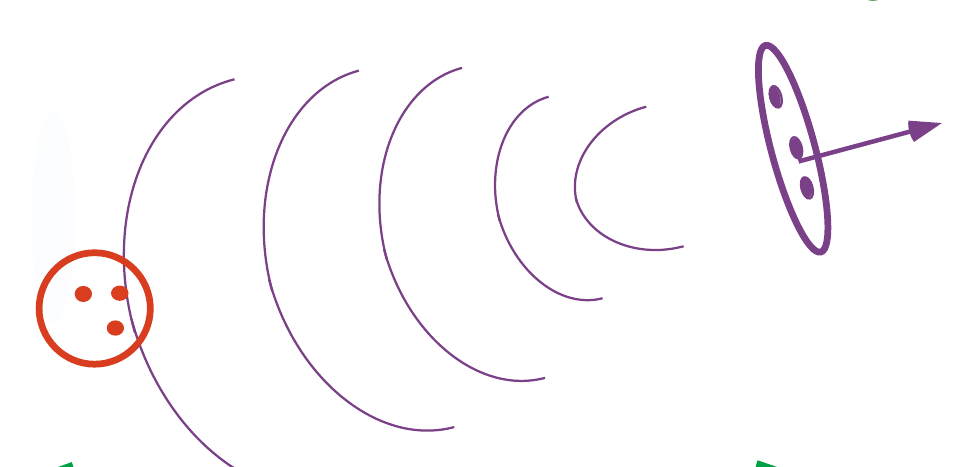}{7 cm} }
\caption{Illustration of separating charges. \label{fig:charges}}
 \end{figure}

We imagine a ``jet-parton", whose coordinate space position we label as $x'$ in its own rest frame, moving with velocity $v$ in the 3-direction toward or away from a ``recoiling parton" of charge $q$, at rest in a system with coordinates $x$.   We can imagine both of these particles as products of the hard scattering.  They may interact with particles in their immediate neighborhood, but here we concentrate on the effect of the recoil parton on the jet parton.    The relevant Lorentz transformation between the two frames is 
\bea
x_3\ =\ \gamma(x'_3\ +\ vt') \ =\  \gamma \Delta\, ,
\eea
which serves to define $\Delta$.   We can think of $x_3'$ as a small, fixed scale, while $x_3$ changes rapidly as the particle recedes, as in Fig.\ \ref{fig:charges}.

  We imagine that the ``collision" was at $\Delta=0$, i.e.\ $t' = -\ \frac{1}vx'_3$, at which time the particles were separated by only a small transverse distance, $x_T=x'_T$, when a large momentum transfer could have taken place between the two.   An estimate of the gauge theory physical effects at later times is found from expressions for the electric field in an abelian theory due to a charge $q$ at rest in the $x$ frame and as seen in the $x'$ frame:
\bea
\quad  E_3'(x') 
\ &=&\  \frac{q\, \gamma\, \Delta} { (x_T^2\ +\ \gamma^2 \Delta^2)^{ 3/2} }\ \sim\  \frac{1}{\gamma^2}\, \frac{q}{\Delta} \, .
 \eea
The force in the 3-direction felt by the parton of charge $q'$ traveling with the jet due to the recoiling charge is just $q'$ times this field.
The electric, $\vec{\bf E}$ field, however, seen by the receding (or approaching) particle is highly contracted, falling off as $1/\gamma^2$ 
 at all times except during an interval whose width decreases as $1/\gamma^2$, and hence as a power of the momentum of the jet.   This suggests that the time development of scattered charges is indeed independent of the hard scattering.

Even in this classical, abelian example, however, the richness of the gauge theory description can be seen.   In contrast to the field strengths,  the vector potential, ${A}^\mu$ is uncontracted, but is mostly a total derivative as seen in the $x'$ frame:
\bea
\hspace{-10mm}
A^\mu(t',x_3') = q \frac{\partial}{\partial x'_\mu}\ \ln  \Delta(t',x'_3)
+
{\cal O}( 1 - \beta) \, .
\label{eq:a-prime}
\eea
 This ``large" part of $A^\mu$  can be removed by a gauge transformation in principle.  The need to implement this freedom makes proofs of factorization challenging in gauge theories \cite{Bodwin:1984hc,Collins:1985ue,Collins:1989gx}.
Nevertheless, apparent non-factorization cancels for inclusive cross sections, and corrections to Eq.\ (\ref{eq:jet-fact-hg}) are of the same order as
the residual ``drag" forces remaining from the vector potential left over from the total derivative.   We can estimate the energy-dependence of these corrections by the relation 
\bea
1- \beta\ \sim\ \frac{1}{2}\; \left [\sqrt{1-\beta^2}\right ]^2\ \sim\ \frac{1}{2}\, \left [ \frac{m}{\sqrt{\hat s}} \right ]^2  \, ,
\eea
with $\sqrt{\hat s}$ the invariant mass of the system made of the jet parton and the recoil parton, assumed to have some mass $m$.
Corrections to the autonomous, {\it i.e.}\ factorized, description of high energy processes in this model are thus power suppressed in momentum transfer, suggesting the size of corrections to factorization.

\begin{figure}
\centerline{\figscale{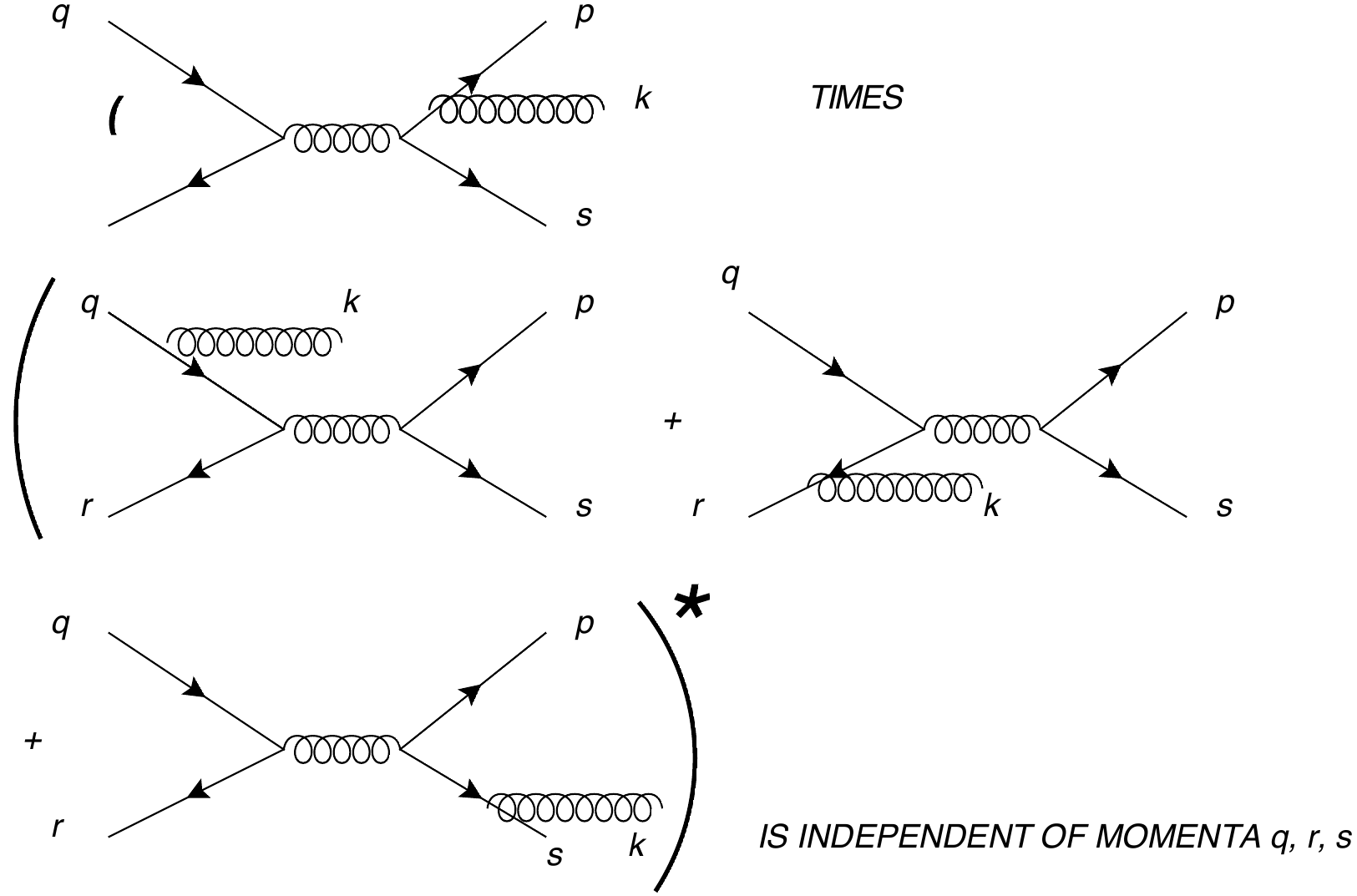}{10cm}}
\caption{In the limit that $k$ is parallel to $p$ the interference between the emission of gluon $k$ from the $p$ line and its emission from lines $q$, $r$ and $s$ is independent of momenta $q$, $r$ and $s$. \label{fig:scalar-g-fact}}
\end{figure}

In QCD, these same features are embedded in Feynman diagrams \cite{Basu:1984ba,Qiu:1990xy}.   When a gluon's momentum,  $k$ becomes collinear to the momentum $p$ of a particle that has emitted it, diagrams that contribute to the amplitude are singular.  This is the quantum field theory analog of the ``uncontracted" classical gauge potential in Eq.\ (\ref{eq:a-prime}). In covariant gauges, these singularities appear in cross sections through interference with emission from other lines in the  collision process.   In general, all diagrams contribute, but in the limit that the gluon is parallel to the particle that emits it, the sum of all diagrams is independent of the other momentum directions, and hence is the same for all hard scattering kinematics (see Fig. \ref{fig:scalar-g-fact}).   This is the mechanism that makes possible the universality of long-distance factors (parton distributions) in hadron-hadron scattering in QCD, and in general it only emerges after a sum over many diagrams.

On a still more technical level, the singular, ``collinear" gluons emitted by a charge carry polarizations that are proportional to their own momentum $(\epsilon^\mu(k) \ \propto \ k^\mu)$, and can give non-factoring contributions that grow with energy in individual  diagrams.  In QCD and other gauge theories,  the gauge invariance of physical quantities ensures that such lines with unphysical polarizations organize themselves into gauge rotations on physical particles \cite{Collins:1989gx,Bauer:2000yr,Becher:2014oda}.   Such gauge rotations are generalizations of phase factors associated with the gauge potential of quantum electrodynamics,
\bea
\Phi_\beta(0)\ =\ P\ \exp \left[\, \int_0^\infty d\lambda\; \beta\cdot A(\lambda \beta^\mu) \, \right]\, ,
\eea
with $P$ path ordering in group space of the nonabelian field $A$, along the semi-infinite lightlike path in the  direction $\beta$.   Ambiguities in the choice of $\beta$ are in many ways analogous to the arbitrariness of the precise factorization scale, described above.

 In the manner of a classical charge moving near the speed of light, the jet only knows the rest of the world as a source of 
unphysically-polarized gluons.   These non-Abelian phases \cite{Yang:1974kj}  or Wilson lines  \cite{Wilson:1974sk} are themselves related to the geometric structure underlying gauge theories (see the commentary on \cite{Yang:1974kj} in \cite{Yang:2005sr}), and appear in numerous contexts beyond the perturbative description on which we've focussed in this talk
\cite{Bialynicki-Birula,Mandelstam:1968hz,Polyakov:1978vu,Susskind:1979up}.   The future will surely see new applications, especially in the transition from infrared safe, perturbative observables to infrared-senstive hadronic observables.   In this context, many results are known to all orders in perturbation theory.    Such results typically point at how perturbation theory transcends itself \cite{'tHooft:1977am,Mueller:1984vh,Sterman:2003wk}, making room for the true long-time behavior of the system, emergent from its gauge interactions.

\section{Conclusions}

Accelerators have confirmed the fundamental degrees of freedom in the gauge theories of the Standard Model directly,  complementing and motivating the great advances in technology that were necessary to probe nature at distance scales down to twenty orders of magnitude below the size of the apparatus.  For the most part, contemporary observations are designed for identifying partonic states, in an effort to detect and reject QCD backgrounds in the search for physics beyond the Standard Model.  

Time will tell whether the gauge theories accessible at accelerators will offer a resolution to dark matter identity, the hierarchy problem, and related mysteries, as they did for once-exotic manifestations of the Standard Model.    I hope so.   Reflecting on the extraordinary range of success for the theories inspired by the gauge concept, sixty years after its extension to nonabelian groups, we may have cause for optimism.

 \section*{Acknowledgments}
 
 I thank the organizers, especially K.K.\ Phua and L.\ Brink, for the invitation and support to participate in the {\it Conference on Sixty Years of Yang-Mills Gauge Theories}, and the Institute of Advanced Studies, Nanyang Technological University for its hospitality.   This work was supported in part by the National Science Foundation under award PHY-1316617.

\end{document}